\documentclass[twocolumn]{aastex6}
\usepackage{graphicx}
\usepackage{epsfig}
\usepackage{amsmath, amssymb , amsthm, fouriernc}
\usepackage{threeparttable}

\newcommand{\src}{2E\,1613.5$-$5053}
\newcommand{\rcw}{RCW\,103}

\newcommand{\swift}{\textit{Swift}}
\newcommand{\chandra}{\textit{Chandra}}

\newcommand{\msun}{\ensuremath{\mathrm{M_\odot}}}

\usepackage{times}

\newcommand{\myemail}{shriharsh@physics.mcgill.ca}
\begin{document}

\title{Near infrared counterpart of 2E\,1613.5$-$5053 the central source in supernova remnant RCW\,103}

\author{ S.~P. Tendulkar, V.~M. Kaspi, R.~F. Archibald, P. Scholz}
\affil{Department of Physics and McGill Space Institute, McGill University, 3600 University St., Montreal QC, H3A 2T8, Canada}
\affil{National Research Council of Canada, Herzberg Astronomy and Astrophysics, Dominion Radio Astrophysical Observatory, P.O. Box 248, Penticton, BC V2A 6J9, Canada}
\email{\myemail}

\begin{abstract}
On 2016 June 22, \src, the puzzling central compact object in supernova remnant \rcw, emitted a magnetar-like burst. Using Director's Discretionary Time, we observed \src\ with the Hubble Space Telescope (WFC3/IR) and we report here on the detection of a previously unseen infrared counterpart. In observations taken on 2016 July 4 and August 11, we detect a new source ($m_\mathrm{F110W} = 26.3\,$AB mag and $m_\mathrm{F160W} = 24.2$\,AB mag) at the \emph{Chandra} position of \src\ which was not detected in HST/NICMOS images from 2002 August 15 and October 8 to a depth of 24.5\,AB mag (F110W) and 25.5\,AB mag (F160W). 
We show that these deep IR observations rule out the possibility of an accreting binary but mimic IR emission properties of magnetars and isolated neutron stars. The presence or absence of a low-mass fallback disk cannot be confirmed from our observations.
\end{abstract}

\section{Introduction}
\label{sec:intro}
\src\ was discovered as a bright X-ray source in the supernova remnant (SNR) RCW\,103 using the \emph{Einstein} X-ray Observatory \citep{tuohy1980}. The nature of \src\ has been mysterious for the past three decades. With soft thermal X-ray emission, an apparent absence of radio detection, a location in the center of an SNR, it was first classified as a Central Compact Object \citep[CCO; ][]{deluca2008a}. However, this classification is fraught with trouble. Unlike CCOs whose X-ray luminosity is usually stable, \src\ shows variations in X-ray luminosity over multiple orders of magnitudes \citep{gotthelf1999, esposito2011} on timescales of months and years.  

\defcitealias{deluca2008b}{dL08}
A suprising 6.67-hr periodicity was discovered with nearly 50\% modulation in the X-ray band \citep{deluca2006} with no hint of faster pulsations. The 6.67-hr periodicity is too slow for the rotation of a young isolated neutron star, requiring exotic explanations for origins or braking mechanisms such as wind and/or disk accretion \citep{deluca2006, li2007}. The periodicity is typical of compact binaries and models of tidal locking with a binary companions \citep{pizzolato2008} and propeller emission from an accretion disk in a pre-low mass X-ray binary \citep{bhadkamkar2009} has been suggested to explain the periodicity as an orbital modulation. However, deep near infrared (NIR) imaging has limited any binary companion to be less massive than an M6 star, too small to support an accretion luminosity of $10^{34-35}\,\mathrm{erg\,s^{-1}}$  \citep[][ hereafter dL08]{deluca2008b}. 

On 2016 June 22, the \swift\ Burst Alert Telescope \citep[BAT,][]{barthelmy2005} detected a millisecond-timescale magnetar-like burst from the region of SNR \rcw\ \citep{dai2016ATel}. \swift\ slewed its X-ray Telescope \citep[XRT,][]{burrows2005} and detected that \src\ was in outburst, with an absorbed 0.5-10\,keV flux of $4 \times 10^{-11}\,\mathrm{erg\,cm^{-2}\,s^{-1}}$ substantially higher than the quiescent absorbed flux of $2 \times 10^{-12}\,\mathrm{erg\,cm^{-2}\,s^{-1}}$ in the same band. The short burst, and the double blackbody + hard power-law (spectral index $\Gamma\approx1.2$) shape of the outburst spectrum support the source being a magnetar \citep{dai2016, rea2016}, but the origin of the 6.67-hr periodicity remains puzzling. The slowing of a magnetar via magnetic field interactions with a fallback disk was suggested and preferred by many authors but the binary scenario has not been completely ruled out \citep{deluca2008b,dai2016,rea2016}. New theoretical work suggests that a neutron star with a high magnetic field ($B\sim5 \times 10^{15}$\,G) and a fallback disk can efficiently decelerate to rotational periods of a few hours. The estimated mass of the fallback disk required to slow down the disk varies from $10^{-9}\,\msun$ \citep{ho2016} to $10^{-5}\,\msun$ \citep{tong2016}. Whether the fallback disk survives the interaction to the present day is an unanswered question.

\subsection{Near IR Counterpart/Companion}
\src\ lies in the Galactic plane ($l = 332^\circ,b = -0.4^\circ$) in a crowded stellar field with high extinction. This makes the identification of the counterpart or companion to \src\ challenging. Previous authors have attempted identification using photometric variability \citep{sanwal2002,mignani2008} and also colors in the NIR \citepalias{deluca2008b}. However, no obvious candidate has stood out from the 7 candidates in or near the 99\% \chandra\ position error ellipse. 

The 2016 June 22 outburst provided an opportunity to look for NIR luminosity variations that have been observed during outbursts of magnetars as well as in accretion binaries. Here we describe our Director's Discretionary Time observations with HST/WFC3 and report a new source that was absent in the 2002 observations.


\section{Observations \& Analysis}
\subsection{2016 Observations}

We requested DDT observations of \src\ in the F160W (\emph{H} band) and F110W (\emph{Y+J} band) filters using the WFC3 instrument. The images were acquired on 2016 July 4 and 2016 August 11, corresponding to 12 and 50 days after the first magnetar-like burst of \src, respectively. The observation details are specified in Table~\ref{tab:hst_observations}. The observations were spaced to detect the likely fading of magnetars over the timescale of a month \citep[see par. ex. ][]{kaspi2014}. However, the average X-ray luminosity of \src\ did not decrease significantly over this time period (see Section~\ref{sec:x-ray}). 

We used the  WFC3/IR camera with a 512$\times$512 pixel ($68\arcsec \times 68\arcsec$) aperture in both filters. We acquired $4 \times 321\,\mathrm{s}$ exposures in the F110W band and $1 \times 105\,\mathrm{s} + 4 \times 321\,\mathrm{s}$ exposures in the F160W band at both epochs. The 321-s exposures in each filter were read out using the \texttt{SPARS25} sampling and the 105-s exposure was acquired using the rapid log-linear \texttt{STEP25} readout to correctly image bright stars in the field\footnote{See the instrument handbook for details: \url{http://www.stsci.edu/hst/wfc3}}. The exposures were dithered with the standard 4 position dither (\texttt{WFC3-IR-DITHER-BOX-MIN}) to improve the sampling of the point spread function (PSF) and to identify and remove cosmic rays. 

We processed the images with the standard \texttt{STSDAS} analysis package in \texttt{IRAF}. We dedistorted and combined the images to a platescale of 55\,mas per pixel in the F160W filter and 37.8\,mas per pixel in the F110W filter using the \texttt{drizzlepac} package. We chose the platescales to sample the point spread function (PSF) in each filter with 2.5 pixels. 

\subsection{2002 Observations}
We downloaded archival NICMOS NIC2 images of \src\ acquired in 2002 August and October from the Space Telescope Archive (Program 9467). The details of the data are specified in Table~\ref{tab:hst_observations}. The F110W images from 2002 were shallow (total exposure of 1870\,s) and only the brightest stars were visible. Hence we only consider the 2002 F110W images to measure upper limits in the analysis.  We did not use the F205W (\emph{K}) band images in this analysis as they were discussed in \citetalias{deluca2008b}. 

The image files from the August and October observations were separately dedistorted and combined using \texttt{MultiDrizzle} to the same plate scales and settings as the WFC3/IR observations.

\subsection{\swift-XRT Observations}
\label{sec:x-ray}
As the IR luminosity of magnetars may vary with the X-ray luminosity, we analyzed the 0.5--10\,keV X-ray data from \swift-XRT observations closest in time to the 2016 HST observations. We created the \swift-XRT spectrum for \src\ from observations 00700791011 (2.1\,ks exposure at 2016 July 4 13:21 UT) and 00030389037 (2.7\,ks exposure at 2016 August 10 01:08 UT) using the automated XRT data analysis tool\footnote{\url{http://www.swift.ac.uk/user\_objects/}} \citep{evans2009}. We fit the spectra with an absorbed blackbody model over 0.5--10.0\,keV using \texttt{XSPEC} v12.9.0n \citep{arnaud1996}. We used the \texttt{wilm} abundance model \citep{wilms2000} and \texttt{bcmc} photoelectric cross sections \citep{balucinskachurch1992}.  The best fit photoelectric column density was $N_\mathrm{H}=1.4\pm0.7\times10^{21}\,\mathrm{cm^{-2}}$.  The measured blackbody temperatures and unabsorbed fluxes were $kT_1 = 0.56\pm0.04$\,keV, $kT_2=0.60\pm0.3$\,keV and $F_{X,1} = 4.5\pm0.4\times10^{-11}\,\mathrm{erg\,cm^{-2}\,s^{-1}}$ and $F_{X,2} = 4.7\pm0.4\times10^{-11}\,\mathrm{erg\,cm^{-2}\,s^{-1}}$, where the subscripts 1 and 2 refer to the July and August epochs, respectively. Thus, we conclude that the X-ray flux did not decrease significantly between our HST observation epochs, consistent with the X-ray light curve reported by \citet{dai2016} and \citet{rea2016}.


\begin{deluxetable}{lclc}
\tablecolumns{4}
\tablecaption{HST Observations of \src. \label{tab:hst_observations}}
\tablewidth{0pt}
\tabletypesize{\footnotesize}
\tablehead{\colhead{Obs ID} & \colhead{Start --- End (UT)} & \colhead{Inst/Filt\tablenotemark{a}} & \colhead{Exp\tablenotemark{b}}}
\startdata
\sidehead{2016 August}
ID4V02VJQ &		2016-08-11 02:48 --- 02:50 & WFC3/F160W    & 105.5 \\
ID4V02020 &		2016-08-11 02:56 --- 04:16 & WFC3/F160W    & 1287\\
ID4V02010 &		2016-08-11 02:50 --- 04:10 & WFC3/F110W    & 1287\\
\sidehead{2016 July}
ID4V01B3Q &		2016-07-04 02:48 --- 02:50 & WFC3/F160W    & 105.5 \\
ID4V01020 &		2016-07-04 02:56 --- 04:16 & WFC3/F160W    & 1287\\
ID4V01010 &		2016-07-04 02:50 --- 04:10 & WFC3/F110W    & 1287\\
\hline
\sidehead{2002 October (archival)}
N8C501010 &		2002-10-08 02:26 --- 04:10 & NIC2/F160W  & 2590.5\\
N8C501020 &		2002-10-08 04:10 --- 06:04 & NIC2/F160W  & 2590.5\\
N8C501030 &		2002-10-08 06:05 --- 09:09 & NIC2/F160W  & 2590.5\\
N8C501040 &		2002-10-08 09:10 --- 11:12 & NIC2/F160W  & 2590.5\\
N8C501050 &		2002-10-08 11:13 --- 12:39 & NIC2/F110W  & 935 \\
\sidehead{2002 August (archival)}
N8C502020 &		2002-08-15 10:28 --- 12:03 & NIC2/F160W  & 2590.5\\
N8C502040 &		2002-08-15 13:40 --- 14:31 & NIC2/F160W  & 2590.5\\
N8C502050 &		2002-08-15 15:19 --- 15:41 & NIC2/F110W  & 935 \\
N8C502030 &		2002-08-15 12:04 --- 12:55 & NIC2/F160W  & 2590.5\\
N8C502010 &		2002-08-15 08:50 --- 10:27 & NIC2/F160W  & 2590.5\\
\enddata
\tablenotetext{a}{Instrument and Filter: WFC3 -- Wide Field Camera 3 / IR, NIC2 -- NICMOS Camera 2.}
\tablenotetext{b}{Exposure time in seconds.}
\end{deluxetable}

\section{Results}
\subsection{PSF Fitting}
For accurate photometry and astrometry, we performed PSF fitting on each image using the \texttt{IDL} code \texttt{StarFinder} \citep{diolaiti2000}. We used a $95 \times 95$\,pixel PSF model (5.2\arcsec\ in F160W and 3.6\arcsec in F110W) to account for the diffraction spikes. We assumed the PSF model to be static over each drizzled image. 

For each image, we used fifteen bright stars in the image to create a model PSF that was fit to all the stars in the image. This step was iterated twice and we ensured that the PSF model was cleaned of contaminating stars. We extracted the pixel coordinates and fluxes of stars in each filter and epoch. We limited the search to sources that were detected with a signal to noise ratio (SNR) greater than 3 and where the normalized PSF fitting correlation was greater than 0.8. The residual images were analysed by eye to verify that no under-fitting or over-fitting had occured. The PSF model created for each image was saved.

We used the 2016 July F160W image as the reference for matching all objects from other images. We first corrected the world coordinate system of the reference image to the 2MASS star positions \citep{skrutskie2006}  using the \texttt{IRAF} task \texttt{ccmap}. The residual fitting error was 0.11\arcsec (root-mean-square). Before fitting, we removed 2MASS sources that corresponded to unresolved stars in the HST images. The positions of stars in other images were matched and transformed to the image coordinate system of 2016 July F160W image using the \texttt{geomap} and \texttt{geoxytran} \texttt{IRAF} tasks. The residuals of the matching were $\approx0.4$\,pix (22\,mas). We matched the detected sources using a search radius of 0.5\,pixels and produced a combined list of sources and fluxes/non-detections. 

\subsection{Photometry}
The flux reported by \texttt{StarFinder} is the integrated flux under the normalized PSF. We converted the flux to AB magnitudes using the \texttt{PHOTFNU} keyword based on the WFC3/IR and NICMOS calibration. While \texttt{StarFinder} reports a formal flux error for each star, this does not account for the error in PSF estimation, PSF variation over the image and the background estimation. We estimated the scatter in fluxes by comparing the fluxes measured in the 2016 July images to 2016 August images and 2002 August images to the 2002 October images. As the image pairs were acquired with the same instrumental configuration separated only by a few months, the scatter in the fluxes should be dominated by the errors arising from the sources discussed above. 

\begin{figure}
  \center
  \includegraphics[width=0.48\textwidth]{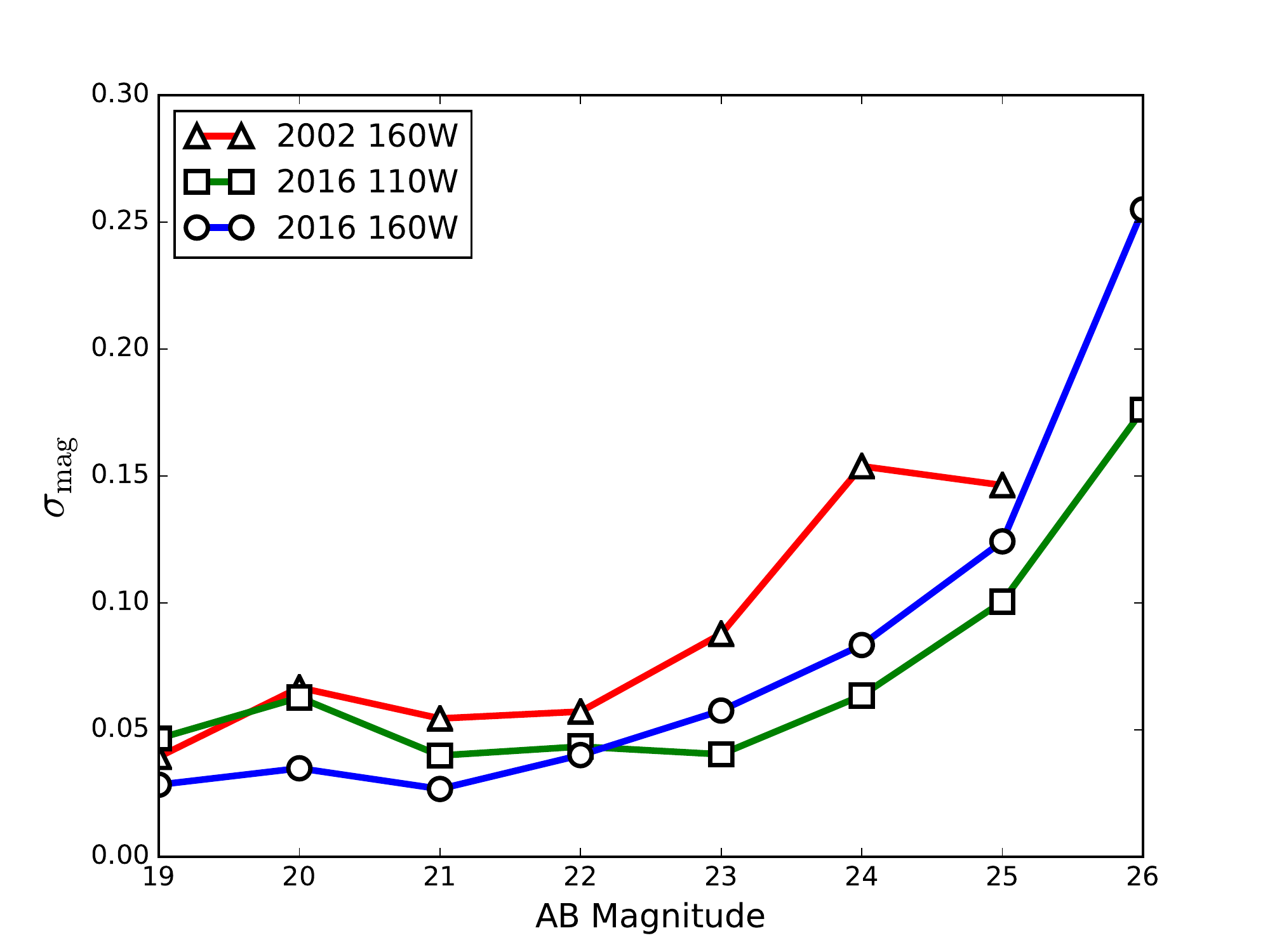}
  \caption{1-$\sigma$ scatter in photometry for the 2016 WFC3 F110W (squares), F160W (circles) and 2002 NICMOS F160W images (triangles). The scatter is due to a combination of PSF modelling errors, PSF variation over the image, background contribution and the Poisson noise. The 2002 NICMOS F110W images did not have sufficiently many stars to accurately estimate the standard deviation in each magnitude bin.}
  \label{fig:mag_err}
\end{figure}

 Figure~\ref{fig:mag_err} shows the measured magnitudes and magnitude differences in the pairs of images and the calculated scatter in 1\,mag bins. The NICMOS F110W images were not used for this analysis as they did not have sufficient stars to accurately estimate the standard deviation in each magnitude bin. Comparing the photometry between WFC3 and NICMOS, we find that the faint star ($>20\,$mag) photometry matches within errors and there is no significant zeropoint difference. The bright star photometry with NICMOS is known to have a non-linearity\footnote{\url{http://www.stsci.edu/hst/nicmos/performance/anomalies/nonlinearity.html}} and it is detectable at a 0.1\,mag level for bright stars.

\subsection{Detected Sources}
\begin{figure}
  \center
  \fbox{\includegraphics[width=0.44\textwidth]{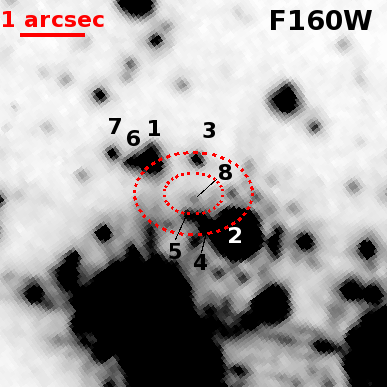}}
  \fbox{\includegraphics[width=0.44\textwidth]{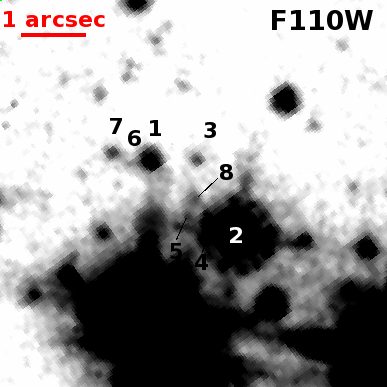}}
  \caption{WFC3 F160W (top panel) and F110W (bottom panel) images of \src\ from July 2016. The stars are labelled as per \citetalias{deluca2008b} and the new detection, source 8 is marked. The dotted ellipse shows the 68\% and 99\% position error ellipse calculated by \citetalias{deluca2008b}.}
  \label{fig:2016images}
\end{figure}

We labeled the sources detected in the field as per the scheme used in \citetalias{deluca2008b} (Figure~\ref{fig:2016images}). We detected a new object, Source 8, in the 2016 images inside the \chandra\ position ellipse. The magnitudes of the source along with photometric scatter (as calculated above) are given in Table~\ref{tab:photometry}.

\begin{figure*}
  \center
  \fbox{\includegraphics[width=0.31\textwidth]{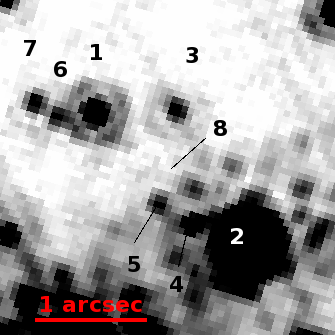}}
  \fbox{\includegraphics[width=0.31\textwidth]{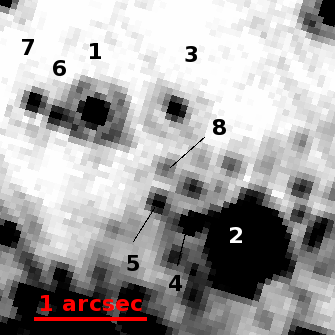}}
  \fbox{\includegraphics[width=0.31\textwidth]{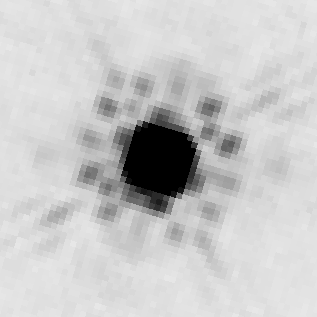}}
  \caption{Verifying the detectability of source 8 in 2002 F160W images. \emph{Left Panel:} Original 02-J-160 image. \emph{Middle Panel:} 02-J-160 image with source 8 injected by scaling the extracted PSF (\emph{right panel}) to an AB magnitude of 24.2\,mag. The source is easily detected. The bright spots to the lower right of source 8 in the middle panel are speckles of the PSF as shown in the right panel.}
  \label{fig:2002images}
\end{figure*}

Source 8 was not detected in the 2002 images (Figure~\ref{fig:2002images}, left panel). As a verification, we converted the measured F160W flux from the 2016 July measurement into the expected NICMOS count rate. Using the PSF extracted from the 2002 August and 2002 October images, we injected a fake source at the location of source 8. The source is clearly visible and also detected by the same analysis pipeline as utilized above (Figure ~\ref{fig:2002images}, middle panel). By reducing the brightness of the injected source till it was not detected in our analysis pipeline, we estimate the limiting brightness of source 8 in 2002 August and 2002 October (Table~\ref{tab:photometry}). We put 3-$\sigma$ upper limits of $m_\mathrm{F110W}>24.5$ and $m_\mathrm{F160W}>25.2$ for individual images. If the 2002 August and October images are combined, the limiting magnitudes are $m_\mathrm{F110W}\gtrsim25$ and $m_\mathrm{F160W}\gtrsim25.5$.

\begin{deluxetable*}{l|cc|cc|cc|cc}
\centering
\tablecolumns{9}
\tablecaption{Photometry of sources near the location of \src.\label{tab:photometry}}
\tablewidth{0pt}
\tabletypesize{\footnotesize}
\tablehead{
  \colhead{} & \multicolumn{4}{c}{2002} & \multicolumn{4}{c}{2016}\\
  \colhead{} & \multicolumn{2}{c}{Aug} & \multicolumn{2}{c}{Oct} & \multicolumn{2}{c}{Jul} & \multicolumn{2}{c}{Aug} \\
  \colhead{\#} & $m_\mathrm{F110W}$ & $m_\mathrm{F160W}$ & $m_\mathrm{F110W}$ & $m_\mathrm{F160W}$ & $m_\mathrm{F110W}$ & $m_\mathrm{F160W}$ & $m_\mathrm{F110W}$ & $m_\mathrm{F160W}$ 
}
\startdata
1 & $23.8\pm0.2$ & $21.20\pm0.06$ & $23.8\pm0.2$ & $21.16\pm0.06$ & $23.84\pm0.07$ & $21.44\pm0.04$ & $23.76\pm0.07$ & $21.40\pm0.04$ \\
2\tablenotemark{a} & $19.1\pm0.1$ & $18.02\pm0.04$ & $19.1\pm0.1$ & $18.01\pm0.04$ & $19.70\pm0.06$ & $18.37\pm0.03$ & $19.63\pm0.06$ & $18.38\pm0.03$ \\
3 & $>24.5$ & $22.92\pm0.09$ & $>24.5$ & $22.91\pm0.09$ & $25.55\pm0.15$ & $23.04\pm0.06$ & $25.45\pm0.15$ & $23.05\pm0.06$ \\
4 & $>24.5$ & $22.50\pm0.08$ & $>24.5$ & $22.61\pm0.08$ & $25.53\pm0.14$ & $22.90\pm0.06$ & $25.65\pm0.15$ & $23.04\pm0.06$ \\
5 & $>24.5$ & $23.19\pm0.11$ & $>24.5$ & $23.20\pm0.11$ & $26.51\pm0.21$ & $23.53\pm0.07$ & $26.19\pm0.20$ & $23.51\pm0.07$ \\
6 & $>24.5$ & $23.25\pm0.11$ & $>24.5$ & $23.20\pm0.11$ & $26.45\pm0.21$ & $23.50\pm0.07$ & $26.26\pm0.20$ & $23.46\pm0.07$ \\
7 & $>24.5$ & $22.96\pm0.09$ & $>24.5$ & $22.94\pm0.09$ & $25.37\pm0.12$ & $23.11\pm0.06$ & $25.26\pm0.12$ & $23.09\pm0.06$\\ 
8\tablenotemark{b} & $>24.5$ & $>25.2       $ & $>24.5$ & $> 25.2      $ & $26.27\pm0.20$ & $24.24\pm0.08$ & $26.39\pm0.21$ & $24.51\pm0.10$ 
\enddata
\tablecomments{All magnitudes are measured in the AB magnitude scale.}
\tablenotetext{a}{Star 2 is affected by the photometric nonlinearity of the NICMOS detector and hence the difference in 2002 and 2016 magnitudes is not astrophysical.}
\tablenotetext{b}{New source detected only in 2016 observations.}
\end{deluxetable*}



\section{Discussion}
\label{sec:discussion}
Following the 2016 magnetar flare and X-ray brightening of \src, we have detected a new infrared source at the X-ray location of \src. The infrared source brightened by at least 1.3\,mag (F160W) compared to the non-detections in previous 2002 observations. Thus, we conclude that this source is associated with \src\ and we discuss the implications, and the physical scenarios for explaining its 6.67-hr X-ray modulation. 

\paragraph{Dust Extinction in IR}
To interpret the nature of the source, we must first try to determine its intrinsic brightness, corrected for extinction. The magnitude of the optical/IR extinction towards \src\ is uncertain. The dust maps of \citet{schlafly2011} estimate $A_V=36$\,mag in the direction of \src\footnote{\url{http://irsa.ipac.caltech.edu/applications/DUST/}}. This is also supported by the average $H-K$ color of the surrounding stars from \citetalias{deluca2008b} who estimate $A_V=20-40$ for the whole field. However, the near-IR spectroscopy of \rcw\ \citep{oliva1989} and the photoelectric absorption column density ($N_\mathrm{H}$) estimated from X-ray observations of \src\ and \rcw\ \citep{foight2016} suggest a much lower value of $A_V=3-6$. 

Here, we discuss both the high extinction case ($A_V=36$\,mag) and the low extinction case ($A_V=3.5$\,mag) assuming the distance of $3.3\,$kpc to \rcw\ \citep{tuohy1980} but we consider that the low extinction value is substantially more likely since it arises from measurements of \src\ and \rcw\ themselves. We also show that the high extinction case leads to infeasible scenarios.

For each case, we discuss whether the IR emission could be due to a companion/accretion disk (binary scenario) in which case the 6.67-hr period could be interpreted as the orbital period. We also discuss, alternatively, whether the IR emission is from the neutron star or a fallback disk (isolated scenario) where the 6.67\,hr period is interpreted as the rotational period of the neutron star.

The 0.5--10\,keV X-ray flux at the 2002 and 2016 observation epochs was approximately $6\times10^{-12}\,\mathrm{erg\,cm^{-2}\,s^{-1}}$ \citepalias{deluca2008b} and $4.5\times10^{-11}\,\mathrm{erg\,cm^{-2}\,s^{-1}}$ \citep{rea2016}, respectively. This corresponds to intrinsic luminosities of $L_X = 7\times10^{33}\,\mathrm{erg\,s^{-1}}$ and $L_X = 5\times10^{34}\,\mathrm{erg\,s^{-1}}$, an increase by a factor of $\sim$7.

\subsection{High Extinction Case}
If $A_V=36\,$mag, the extinction in the F110W and F160W bands is 9.5\,mag and 6.0\,mag, respectively. Thus, including a distance modulus of $5\log(3.3\,\mathrm{kpc}/10\,\mathrm{pc})=12.6$\,mag, the absolute AB magnitudes of source 8 in 2016 are $4.2$  (F110W) and $5.6$ (F160W). The corresponding limits in 2002 are  $>2.4$, and $>6.9$\,AB mag, respectively.

We compared the absolute magnitudes to stellar spectrophotometry \citep[][Table 15.7]{allens4thed} and white dwarf models \citep[][ and references therein\footnote{\url{http://www.astro.umontreal.ca/~bergeron/CoolingModels}}]{tremblay2011, bergeron2011}. The 2002 NICMOS upper limits are consistent with main-sequence stars cooler than M2 or with DA and DB white dwarfs as companions to \src. However, \citetalias{deluca2008b} used deeper $K_s$ band VLT upper limits to rule out any binary companions brighter than a M6-M8 dwarf. Also, the F110W$-$F160W color of $-1.4$ in the 2016 observations is bluer than blackbodies of $10^{15}$\,K, ruling out any interpretation of the infrared flux as blackbody emission from a star or an accretion disk. Interpreting the color as power-law emission ($\nu^{\alpha}$), we get $\alpha\approx 4.5$, rising far steeply than the observed spectra from low mass X-ray binary accretion disks ($0.5\lesssim\alpha \lesssim1.5$) \citep{hynes2005}.

The 2016 IR luminosity of \src\ corresponds to $L_\mathrm{F110W}=10^{33}\,\mathrm{erg\,s^{-1}}$ and $L_\mathrm{F160W}=2\times10^{32}\,\mathrm{erg\,s^{-1}}$. Comparing the IR and X-ray luminosities, we get $L_X/L_{F110W,F160W} \approx 50-200$. This ratio of X-ray to IR luminosities is also significantly lower than the values of $10^4$ observed for isolated neutron stars, magnetars and CCOs \citep{fesen2006,wang2006,mignani2008}.

Thus, we find that the high extinction scenario leads to astrophysically infeasible cases and we do not discuss it further.

\subsection{Low Extinction Case}
If $A_V=3.6$, the extinction in the F110W and F160W bands is 0.9\,mag and 0.6\,mag, respectively. This leads to absolute AB magnitudes in 2016 of 12.8 (F110W) and 11.0 (F160W). The corresponding 2002 upper limits are $>11.0$ AB mag and $>12.3$\,AB mag, respectively.
 
For a companion object, the 2002 upper limits are inconsistent with the coolest M-dwarfs. The absolute magnitudes of very compact cooler white dwarfs are consistent with this non-detection. However, a white dwarf companion in an LXMB must necessarily form after the neutron star and hence must be younger than the neutron star or the supernova remnant. Among white dwarfs younger than 2\,kyr, only the compact hydrogen atmosphere (DA) white dwarfs (surface gravity $\log g \gtrsim 9.5$) are consistent with the F160W and F110W upper limits from 2002. The high $\log g$ implies a mass of $\sim1.3\,\mathrm{M_\odot}$ for C, O and mixed CO cores \citep{fontaine2001} and a very tiny radius of $\lesssim5000\,$km. This radius is much smaller than the orbital radius ($1.7\times10^{6}\,$km) for a 1.3\,$\mathrm{M_\odot}$ white dwarf orbiting with a 1.4\,$\mathrm{M_\odot}$ neutron star with a 6.67-hr period and the corresponding Roche lobe radius. Hence such a white dwarf could not provide the accretion power for the X-ray luminosity.

The F110W and F160W luminosities are  $L_\mathrm{F110W}=4\times 10^{29}\,\mathrm{erg\,s^{-1}}$ and $L_\mathrm{F160W}=7\times10^{29}\,\mathrm{erg\,s^{-1}}$. The corresponding X-ray to IR fluence ratios $L_X/L_\mathrm{F110W,F160W} \approx 10^{5}$ are consistent with those of magnetars such as 4U\,0142+61 \citep{hulleman2004}, 1E\,1048.1$-$5937 \citep{tam2008} and limits on other magnetars \citep{fesen2006,wang2006,mignani2008}. It is not clear, however, whether this emission arises from the magnetosphere of the neutron star or whether it arises from a fallback disk, as has been suggested around 4U\,0142+61 \citep{wang2006} and 1E\,2259+586 \citep{kaplan2009}. \citet{wang2007a} measured \emph{Spitzer} flux upper limits to be $10^{-4}\,$Jy (4.5\,$\mu$m) and $3\times10^{-4}\,$Jy (8\,$\mu$m). These measurements do not rule out the presence of a disk as massive as the one around 4U\,0142+61 (10\,$\mathrm{M_\oplus} = 6 \times 10^{28}\,$g). Indeed, the amount of material required to slow down the magnetar to its current period is tiny --- \citet{ho2016} estimate it to be $10^{24}$\,g while \citet{tong2016} estimate the mass to be $10^{28}\,$g. 

Thus, while the presence or absence of a fallback disk cannot be confirmed at this point, we have shown that the binary scenarios for the evolution of \src\ can be ruled out with a high level of confidence. Further understanding of the nature of \src\ can be achieved via spectroscopy of the faint IR source to search for disk emission features and whether the continuum is better described by a power law spectrum or a disk blackbody spectrum. While this is extremely challenging with current observational capabilities, it may be possible with the James Webb Space Telescope. 

\acknowledgements
The authors thank the Hubble Space Telescope operations teams for their speed and flexibility scheduling these observations.
  
Based on observations made with the NASA/ESA Hubble Space Telescope, obtained from the Data Archive at the Space Telescope Science Institute, which is operated by the Association of Universities for Research in Astronomy, Inc., under NASA contract NAS 5-26555. These observations are associated with programs \#9467 and \#14814. This work also made use of data supplied by the UK Swift Science Data Centre at the University of Leicester.

 S.P.T acknowledges support from a McGill Astrophysics postdoctoral fellowship. V.M.K. receives support from an NSERC Discovery Grant, an Accelerator Supplement and from the Gerhard Herzberg Award, an R. Howard Webster Foundation Fellowship from the Canadian Institute for Advanced Study, the Canada Research Chairs Program, and the Lorne Trottier Chair in Astrophysics and Cosmology. R.F.A. acknowledges support from an  NSERC CGSD. P.S. received support from a Schulich Graduate Fellowship from McGill University and holds a Covington Fellowship at DRAO.

\bibliographystyle{aasjournal}
\bibliography{paper}

\end{document}